# Isoscaling in Reactions on Enriched Tin Isotopes and Nuclear Symmetry Energy

A.S.Danagulyan[a], A.R.Balabekyan[a], G.H.Hovhannisyan[a], N.A.Demekhina[b],
J.Adam[c,d], V.G.Kalinnikov[c], M.I.Krivopustov[c], V.M.Tsoupko -Sitnikov[c]

[a]*Yerevan State University, Alex Manoogian 1, 0025, Armenia*
[b]*Yerevan Physics Institute, 0025, Armenia*
[c]*Joint Institute for Nuclear Rrserch, Dubna 141980, Russia*
[d]*The Institute of Nuclear Physics, Czech Republic*

Corresponding author: A.S.Danagulyan danag@ysu.am

**Abstract**

Isospin effects in $^{12}$C ion induced reactions on enriched tin isotopes are investigated. The isoscaling parameter B is obtained for different mass regions of product nuclei. It is shown that the isoscaling parameter is sensitive to the formation mechanism of products, and increases as the difference in the asymmetry is increasing. Using the excitation energy obtained with a catcher technique temperatures, density ratio $\rho/\rho_0$ for product nuclei in different mass regions and the values of symmetry energy coefficient for light mass regions of product nuclei for proton and deuteron induced reactions are determined.

*PACS*: 25.70.Pq

*Keywords:* Ion beam, isoscaling, multifragmentation, symmetry energy, nuclear equation of state, asymmetry

**1. Introduction**

In the past few years there is a growing interest of isospin effects in nuclear reactions, which are qualified as "isoscaling" [1-10] in the research papers.

The investigations of the relation between the entrance channel isospin and isotopic distribution of reaction products are receiving increasing attention because of their importance in the modern problem of extraction of information on the symmetry term in the nuclear equation-of state (EOS) in regions away from normal density under laboratory controlled conditions. Asymmetric nuclear matter models at high density have been tested so far only in astrophysics contexts, in particular, in supernovae explosions and neutron stars. The investigation of multifragmentation allows one to obtain some information about charge equilibration and provides stringent tests for reaction models.

It was observed that in many different types of heavy-ion reactions [1-5, 8, 9] the ratios of light–isotope yields from neutron-rich and neutron-poor systems follow the law of isoscaling.

$$R_{21} = Y_2(N,Z)/Y_1(N,Z) = C \exp(\alpha N + \beta Z) \qquad (1)$$

where $Y(N,Z)$ is the yield of a fragment with Z protons and N neutrons. Indices "1" and "2" correspond to different targets with different isotopic compositions, with "2" corresponding to more neutron-rich target. C is a normalization coefficient. The parameters $\alpha$ and $\beta$ were expressed using the difference of chemical potentials of the two systems as follows: $\alpha = \Delta\mu_n/T$, and $\beta = \Delta\mu_p/T$ [4], where T is the temperature of the excited nucleus.

Recently experimental and theoretical physicists have focused their attention on the investigation of the formation mechanisms of heavy fragments ($Z \geq 20$) by different projectiles ($\gamma$ rays, $\pi^-$ meson, etc.). Experiments with a direct detection of heavy fragments do not provide a comprehensive understanding of fragments in this mass region. The induced-radioactivity method adds more possibilities so that the investigation of mechanisms of heavy-fragment production becomes more realistic [11,12].

In our previous studies of the nuclear reactions induced by protons with energies of 0.66 and 8.1 GeV on targets of tin isotopes the isoscaling behavior was shown using the induced activity method [12] and was described in terms of the third component of the fragment isospin $t_3=(N-Z)/2$. The isotopic ratio of the product formation cross sections has been considered in the following form:

$$R_{21}=Y_2(N,Z)/Y_1(N,Z)=\exp(C+Bt_3) \qquad (2)$$

where C and B are fitting parameters. The parameter B is related to the difference of the chemical potentials of protons and neutrons in the fragment and depends on the temperature of the excited nucleus ($B=2\alpha$, where $\alpha$ is a coefficient in formula (1)). A similar form of description of isotopic dependency of the yields of products in the reactions initiated by light ions with energies of 0.66 GeV to 15.3 GeV on the separated tin isotopes was used in [6]. Our last studies [13] show the dependence of values of parameter B on mass regions of the product nuclei.

In this work we are studying isotopic effects of light, medium-mass and heavy nuclei production in reactions induced by $^{12}$C ions with the energy of 2.2 GeV/N on enriched tin isotopes ($^{112,118,120,124}$Sn).

## 2. Experimental details

The targets of tin isotopes $^{112,118,120,124}$Sn (enrichments - 92.6%, 98.7%, 99.6%, 95.9%, respectively) were irradiated at the Nuclotron of the JINR (Joint Institute for Nuclear Research at Dubna) by $^{12}$C ion beam with the energy of 2,2 GeV/N. For beam monitoring we employed the reaction $^{27}$Al ($^{12}$C,X) $^{24}$Na, which cross-section is 19,4 mb at the energy of 2,1 GeV/N [14]. The target Al foils were of the same size as our targets. The targets consisted of high-purity metal foils with thickness of 390mg/cm$^2$, 61mg/cm$^2$, 70mg/cm$^2$ and 72,7mg/cm$^2$, respectively. The $\gamma$-activity induced in the targets was measured by HpGe detectors. The radioactive nuclei were identified by their half-lives and by the characteristic $\gamma$-lines. The obtained spectra were processed by the software package DEIMOS. Absolute cross-sections of about 105 residuals were obtained from each target.

## 3. Isotopic effect

The discussions of experimental values of cross section ratios affirmed the existence of isoeffect in nuclear reactions by $^{12}$C ion beams with separated tin isotopes. The experimental scaling $R_{21}=Y_2(N,Z)/Y_1(N,Z)=\exp(C+Bt_3)$ of yield ratios with the third component of product isospin is observed in the all reaction channels. The dependence of $R_{21}= Y_2/Y_1$ on $t_3$ for different mass-regions of product nuclei induced by $^{12}$C ion beam ($\Delta N_t=12$) is shown in Fig. 1.

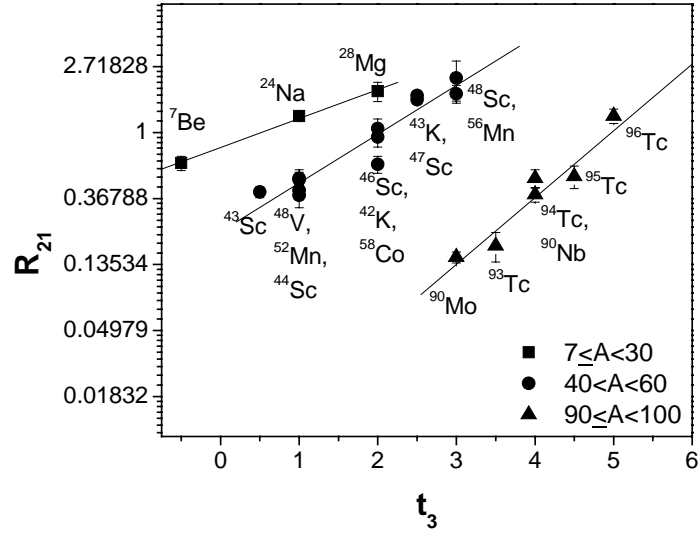

Figure 1. Yield ratio $R_{21}$ versus the third component of the fragment isospin $t_3$. $R_{21}$ is the ratio of product yields from reactions $^{12}C+^{124}Sn$ and $^{12}C+^{112}Sn$. The lines are the result of exponential fits according to Eq. (2) for different mass regions.

The values of fitting parameter B obtained from experimental data are given in Table 1 and Table 2 for different mass regions for target pairs $^{124}Sn/^{112}Sn$ ($\Delta N_t=12$, $(Z_1^2/A_1^2-Z_2^2/A_2^2)=0.03675$), $^{124}Sn/^{118}Sn$ ($\Delta N_t=6$, $(Z_1^2/A_1^2-Z_2^2/A_2^2)=0.01695$), $^{120}Sn/^{112}Sn$ ($\Delta N_t=8$, $(Z_1^2/A_1^2-Z_2^2/A_2^2)=0.02569$), $^{124}Sn/^{120}Sn$ ($\Delta N_t=4$, $(Z_1^2/A_1^2-Z_2^2/A_2^2)=0.01102$), where $N_t$ is the neutron number of the target, $Z_i/A_i$ is the asymmetry of the equilibrated emitting source. For getting precision data for the fitting parameter B mass regions of middle-mass product nuclei (40<A<60, 80<A<90, 90<A<100) were concretized and the statistics was enriched for the mentioned regions by including data for new product nuclei. As a consequence, the new data of B for 8.1GeV proton induced reactions differ from the old ones, while for 3.65A GeV initial energy they are in agreement with data in [13] within errors. We use values of parameter B obtained in our earlier works [13,15,16] for deuteron-induced reactions and different energy proton-induced reactions for comparison. These values are also given in the Tables for the same mass regions as for $^{12}C$ projectile. All product nuclei, which are used for fittings in different mass regions, are brought in Table 3. For each projectile we used only those of them, which experimental cross sections from the corresponding reaction had been determined.

Table 1.
Mean values of the fitting parameter B for different projectiles and target pairs $^{124}Sn/^{112}Sn$ ($\Delta N=12$), $^{120}Sn/^{112}Sn$ ($\Delta N_t=8$).

| Mass region of the product nuclei | protons 0.66 GeV $^{124}Sn/^{112}Sn$ ($\Delta N_t=12$) | protons 3.65 GeV $^{124}Sn/^{112}Sn$ ($\Delta N_t=12$) | deuterons 3.65 GeV/N $^{124}Sn/^{112}Sn$ ($\Delta N_t=12$) | protons 8.1 GeV $^{124}Sn/^{112}Sn$ ($\Delta N_t=12$) | $^{12}C$ 2.2GeV/N $^{124}Sn/^{112}Sn$ ($\Delta N_t=12$) | $^{12}C$ 2.2GeV/N $^{120}Sn/^{112}Sn$ ($\Delta N_t=8$) |
|---|---|---|---|---|---|---|
| 7≤A<30 | – | 0.57±0.02 | 0.55±0.07 | 0.57±0.01 | 0.44±0.03 | 0.26±0.12 |
| 40<A<60 | – | 0.74±0.09 | 0.73±0.13 | – | 0.74±0.07 | 0.55±0.05 |
| 60<A<80 | 1.28±0.39 | 0.83±0.14 | 0.80±0.11 | 0.75±0.10 | 0.88±0.27 | 0.61±.0.15 |
| 80<A<90 | 1.05±0.21 | 0.85±0.15 | 0.87±0.20 | 0.88±0.16 | 0.88±0.22 | 0.70±0.17 |
| 90≤A<100 | 1.05±0.06 | 0.88±0.15 | 0.90±0.20 | 1.00±0.23 | 1.01±0.24 | 0.70±0.12 |
| 100≤A<110 | 1.22±0.10 | 1.33±0.19 | 0.98±0.19 | 1.36±0.24 | 1.23±0.27 | 0.87±0.17 |

Table 2.
Mean values of the fitting parameter B for different projectiles and target pairs $^{124}Sn/^{118}Sn$ ($\Delta N=6$), $^{124}Sn/^{120}Sn$ ($\Delta N_t=4$).

| Mass region of the product nuclei | protons 0.66 GeV $^{124}Sn/^{118}Sn$ ($\Delta N_t=6$) | protons 3.65 GeV $^{124}Sn/^{118}Sn$ ($\Delta N_t=6$) | deuterons 3.65 GeV/N $^{124}Sn/^{118}Sn$ ($\Delta N_t=6$) | protons 8.1 GeV $^{124}Sn/^{118}Sn$ ($\Delta N_t=6$) | $^{12}C$ 2.2GeV/N $^{124}Sn/^{118}Sn$ ($\Delta N_t=6$) | $^{12}C$ 2.2GeV/N $^{124}Sn/^{120}Sn$ ($\Delta N_t=4$) |
|---|---|---|---|---|---|---|
| 7≤A<30 | – | 0.29±0.01 | 0.26±0.05 | 0.34±0.03 | 0.19±0.04 | 0.18±0.09 |
| 40<A<60 | – | 0.41±0.04 | 0.38±0.04 | 0.46±0.09 | 0.44±0.05 | 0.22±0.04 |
| 60<A<80 | 0.58±0.15 | 0.42±0.07 | 0.42±0.09 | 0.47±0.05 | 0.47±0.08 | 0.32±.0.10 |
| 80<A<90 | 0.47±0.17 | 0.44±0.12 | 0.41±0.13 | 0.50±0.18 | 0.48±0.07 | 0.34±0.07 |
| 100≤A<110 | 0.60±0.06 | 0.54±0.07 | 0.49±0.07 | – | 0.59±0.13 | 0.40±0.09 |

Table 3.
Elements included in corresponding mass regions for fitting by Eq. (2).

| Mass region of the product nuclei | Product nuclei used for fitting by Eq. (2) |
|---|---|
| 7≤A<30 | $^{7}$Be, $^{22}$Na, $^{24}$Na, $^{28}$Mg |
| 40<A<60 | $^{42,43}$K, $^{43,44,46,48}$Sc, $^{48}$V, $^{52,54,56}$Mn, $^{56,57,58}$Co |
| 60<A<80 | $^{66,67}$Ga, $^{69}$Ge, $^{70,17,72,74,76}$As, $^{73,75}$Se, $^{75,76,77}$Br, $^{76,77,79}$Kr |
| 80<A<90 | $^{86,88,89}$Zr, $^{82,83}$Sr, $^{81,82,83,84}$Rb, $^{84,86,87}$Y |
| 90≤A<100 | $^{93,94,95,96}$Tc, $^{94,95,97}$Ru, $^{90,93}$Mo, $^{90,96}$Nb |
| 100≤A<110 | $^{103,105,106,110}$Ag, $^{100,101,105}$Rh, $^{100,101}$Pd, $^{100,103}$Ru |

As we can see from Tables 1 and 2, parameter B namely increases with increase of the mass number of product nuclei in all sorts of projectiles and for all values of $\Delta N_t$. Therefore, parameter B is very sensitive to formation mechanism of the corresponding products.

The low mean value of the parameter B (see Tables 1 and 2) for light fragments (7≤A<30) can be understood as an evidence of multifragmentation conditions [3,4,12]. In the production of intermediate mass fragments (A=40÷80) the spallation mechanism obviously participates, but at the same time the multifragmentation contribution is also present, while heavy residuals (near of target mass) are formed via spallation only [3]. This characteristic dependence of parameter B on masses of the products is observed for all projectiles except for protons with the energy of 0.66 GeV (see Table1). Nuclei in the mass region 40<A<60 are not formed in any tin targets, though the intensity of the 0.66 GeV energy proton beam on Phasotron was $10^3$ times larger than that for other projectiles on Nuclotron or Synchrophasotron [15]. It is possible that at comparatively low energies of projectiles a spallation mechanism, and not multifragmentation takes place in the formation of the product nuclei and the cross-sections calculated by INC model for these nuclei are very small.

In our previous works [15,16] for 8.1GeV proton induced reactions besides the experimental values of cross sections of product nuclei those calculated with the Cascade Evaporation and SMM models have also been given. We analyze these calculated values in frame of isoeffect. For light nuclei and nuclei in mass region of 40<A<60 the cross sections within the Cascade Evaporation model are smaller (about 10-30 times) than the experimental ones. Therefore, the selected model cannot describe the formation of product nuclei in these mass regions. However, the calculations for nuclei in 60<A<100 mass region are close to the experimental values of cross sections, except for a few cases. This means that product nuclei close to the target are formed in cascade-evaporation process. Isoeffect is observed in this mass region, and values of fitting parameter B from model calculations are given in Table 4.

Discussion of SMM model calculations shows that it describes the production of light and 40<A<60 mass nuclei rather well, with a few exceptions. For the above-mentioned mass regions isoscaling is also observed and the determined values of fitting parameter B are also brought in Table 4.

Calculations with LAQGSM model [13] for 3.65A GeV proton and deuteron induced reactions are also brought in Table4. The Los Alamos version of the Quark-Gluon-String Model describes reactions as a three-stage process: IntraNuclear Cascade (INC), followed by preequilibrium emission of particles during the equilibration of excited residual nuclei formed during the INC, followed by evaporation of particles. As it can be seen from [13 fig.4, 5] the LAQGSM model doesn't reflect the experimental values of cross sections of light product nuclei, for what calculations of parameter B are absent for this mass region. For the other mass regions isoscaling is observed and values of parameter B are given in Table 4.

Table 4. The calculated values of parameter B ($\Delta N_t=12$) obtained from different models.

| Mass region of the product nuclei | Proton 8.1GeV SMM | Proton 8.1GeV INC | Proton 3.65GeV LAQGSM | Deuteron 3.65GeV/N LAQGSM |
|---|---|---|---|---|
| 7≤A<30 | 0.93±0.19 | - | - | - |
| 40<A<60 | 0.93±0.07 | - | 0.59±0.03 | 0.51±0.04 |
| 60<A<80 | 1.06±0.08 | 0.86±0.16 | 0.78±0.12 | 0.85±0.09 |
| 80<A<90 | - | 1.47±0.29 | 1.16±0.19 | 1.06±0.26 |
| 90≤A<100 | - | 1.66±0.05 | 1.31±0.076 | 1.08±0.15 |

From Table 4 we can see the increase of B with increase of the mean mass region. The comparison shows that the values of parameter B calculated with SMM and Cascade Evaporation models are larger (about 1.5-2 times) than the experimental ones (Tab.1), while the calculations with LAQGSM model are closer to the experimental values within errors. Further detailed theoretical investigation is needed to properly understanding this phenomenon.

The experimental values of parameter B weakly decreases with increasing the projectile energy for light product nuclei (see Tab.1). This dependence for target pair $^{124}Sn/^{112}Sn$ is shown in Fig.2. A stronger energetic dependence of parameter $\beta_{t3}$ for more light nuclei is observed in [6]. And as it is evident from Tables 1,2 energetic dependence of parameter B is absent for middle and heavier mass product nuclei.

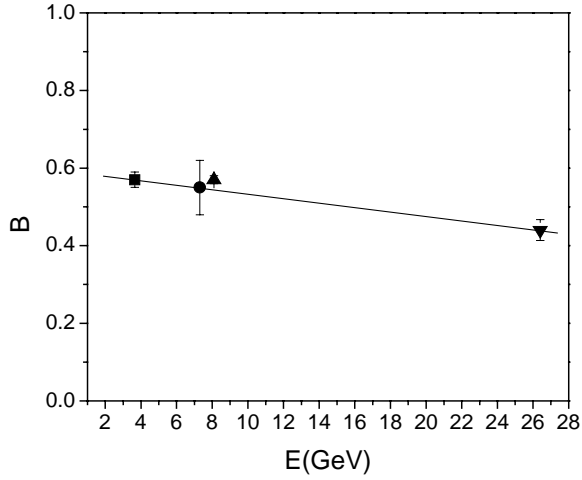

Figure 2. The projectile energy dependence of the parameter B of light product nuclei (7≤A<30) for target pair $^{124}$Sn/$^{112}$Sn. Experimental points for projectiles: ■-proton (3.65GeV), ●-deuteron (7.3GeV), ▲-proton (8.1GeV), ▼-$^{12}$C (26.4GeV). The solid line is the fit of the data of the current work.

Further studies have shown that for reactions with $^{12}$C projectile the parameter B depends on the target neutron excess ($\Delta N_t$) or the difference of asymmetry according to

$$B = c + d \cdot [(Z_t/A_{1t})^2 - (Z_t/A_{2t})^2] \qquad (3)$$

This dependence is shown in Fig.3. Experimental values of B for target pairs $^{124}$Sn/$^{120}$Sn, $^{124}$Sn/$^{118}$Sn, $^{120}$Sn/$^{112}$Sn and $^{124}$Sn/$^{112}$Sn have been used for this fitting. Since for the same target pair $\Delta N_t=0$, the difference of chemical potentials is also 0 ($\Delta\mu=0$), we used zero differences of asymmetry in fitting.

For light nuclei (see Fig.3) the slope ($d=11.09\pm1.33$) is smaller than that for nuclei close to the target 100<A<110 ($d=33.30\pm0.75$).

As it is shown in Figure 3, the parameter $B=2\alpha$ increases as the difference in the asymmetry is increasing. A similar behavior of parameter $B=2\alpha$ (from $\Delta N_t$) we have obtained in proton and deuteron induced reactions with energy of 3,65A GeV on the same targets [13]. This dependence is consistent with the Antisymmetrized Molecular-Dynamics model (AMD) [18], which predicts an increase of isoscaling parameters α and β (see formula (1)) with increasing the differences in the asymmetry of the two considered reactions.

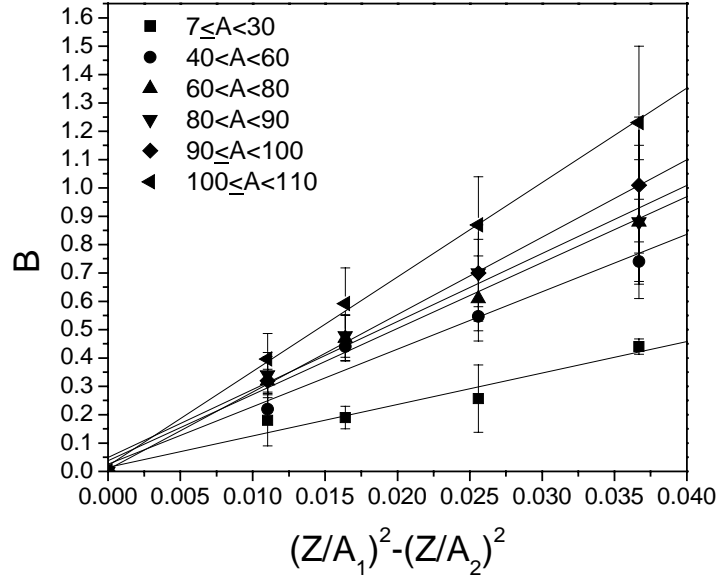

Figure 3. The isoscaling parameter B versus asymmetry for $^{12}$C+Sn reactions. Lines are the fitting results to the data according to Eq.(3). The fitting parameters are: $d$=11.09±1.33, $c$=0.014±0.029 for mass region 7≤A<30; $d$=20.27±1.97, $c$=0.026±0.043 for 40<A<60; $d$=23.3±1.51, $c$=0.038±0.033 for 60<A<90; $d$=27.32±0.39, $c$=0.0067±0.0089 for 90≤A<100 and $d$=33.30±0.75, $c$=0.0198±0.0163 for 100≤A<110. The values of B from Table 1 and 2 are used.

### 4. Isoscaling, excitation energy, temperature, symmetry energy

Investigation of isoeffect allows us to obtain the symmetry energy coefficient γ in the symmetry term in the equation of state. In conditions of chemical equilibrium there are the following relations between the isoscaling parameter α (B=2α), the difference of the chemical potentials of nucleons (Δμ), the temperature T of emitting source, the symmetry energy and symmetry energy coefficient γ [4,6,18]:

$$\alpha = \Delta\mu / T \tag{4}$$

$$E_{sym}(A,Z) = \gamma (A - 2Z)^2 / A \tag{4a}$$

$$\alpha T = \Delta\mu_n = \mu_{n2} - \mu_{n1} = 4\gamma(Z_1^2/A_1^2 - Z_2^2/A_2^2) \tag{4b}$$

where $Z_1$, $A_1$ and $Z_2$, $A_2$ are the charges and mass numbers of the two fragmenting systems after cascade. The ratio $Z_i^2/A_i^2$ is the same as that for the target [5]. The temperature T is generally obtained from the double ratios of cross sections of two light isotopes in multifragmentation process, such as $^{3,4}$He / $^{6,7}$Li , $^{4,6}$He / $^{6,8}$Li , $^{6,8}$He / $^{6,8}$Li double ratios from p+Sn$^{112}$ and p+Sn$^{124}$ reactions [6]. In the present work we use the expanding Fermi gas model predictions according to which a dependence of temperature T on the excitation energy of emitting source [19] (and references therein) is the following:

$$T = [K_0 \, (\rho/\rho_0)^{0.69} \, E^*]^{0.5} \tag{5}$$

where $K_0 = 10$ is the inversed level density parameter. The value of degree for density ratio is taken from [19-21].

In our last work [22] we determine the recoil properties of nuclei with "thick-target thick-catcher" experiment using the induced activity method. Values of excitation energies of about 30 product nuclei formed in $p+^{118}Sn$ and $d+^{118}Sn$ reactions are determined [22]. This allows us to determine $\rho/\rho_0$ using the excitation energy ($E^*/A_{res}$) dependence curve of $\rho/\rho_0$ [19]. The indicated dependence is obtained for $^{58}Fe+^{58}Fe$, $^{58}Fe+^{58}Ni$ and $^{58}Ni+^{58}Ni$ reactions (A=116 for the system) at beam energies of 30, 40 and 47 A MeV, where the experimental points are in agreement with SMM calculations [19]. In the present work a mean excitation energy for light isotopes $Na^{24}$ and $Mg^{28}$ is 591,5 MeV [22]. Mean excitation energy per nucleon is $\bar{E}^*/A_{res} \approx 5.58$ MeV, where $A_{res} \approx 106$ for the $^{118}Sn$ target, according to INC calculations. $A_{res}$ is the residual mass number of the system after cascade process (the multiplicity of the emitted particles at $E_p=3.65$ GeV is $\Delta A=n_p+n_n=13$, $A_{res}=A_t+1(2)-\Delta A$) [23]. The density ratio $\rho/\rho_0 = 0.5$ is deduced for $\bar{E}^*/A_{res} \approx 5.58$ MeV [19], which indicates on liquid-gas transition process. This result agrees with that in Antisymmetrized Molecular Dynamics (AMD) model [18]. Values of $\rho/\rho_0$ are also determined for all the other mass regions with the above-mentioned method for proton and deuteron induced reactions.

The temperature T is calculated from Eq. (5). Having the values of $\alpha$ and T, the symmetry coefficient $\gamma$ is determined from Eq. 4, 4a, and 4b for the light mass regions only, because the light fragments are formed in the multifragmentation process. $\gamma$ equals to 11.45 MeV and 10.48 MeV in proton and deuteron induced reactions, respectively. It should be mentioned that these calculations are done for products after secondary emission. Undoubtedly, the values of the symmetry coefficient will slightly increase for primary processes with about 5-8% according to [4] or 20% according to [6]. Errors of $\gamma$ are about 15-20%, provided by the errors of the forward velocity $\upsilon$ and $E^*$ calculations with Two-Step Vector model [22], by $\rho/\rho_0$ calculations from SMM model and the experimental data [19]. The obtained values of $\gamma$ are smaller than the adopted standard value of $\gamma = 23 \div 25$ MeV at normal density.

The obtained data of excitation energy per nucleon, $\alpha$, $\rho/\rho_0$ and T are brought in Table 5 and Table 6 for proton and deuteron induced reactions, respectively.

Table 5
Average values of $\bar{E}^*/A_{res}$, $\alpha$, $\rho/\rho_0$, T for product nuclei produced in proton induced reaction with 3.65 GeV incident energy. The values of $\alpha=B/2$ are taken from Tab.1

| *Mass region* | $\bar{E}^*/A_{res}$ | $\alpha=B/2$ | $\rho/\rho_0$ | *T (MeV)* |
|---|---|---|---|---|
| 24≤A≤28 | 5.58 | 0.285 | 0.50 | 5.90 |
| 42≤A≤58 | 3.86 | 0.37 | 0.70 | 5.49 |
| 67≤A≤77 | 2.89 | 0.415 | 0.80 | 4.97 |
| 81≤A≤89 | 2.72 | 0.425 | 0.85 | 4.93 |
| 90≤A≤99 | 1.70 | 0.44 | 0.98 | 4.10 |
| 104≤A≤109 | 0.77 | 0.665 | 1 | 2.77 |

Table 6
Average values of Ē*/A$_{res}$, α, ρ/ρ$_0$, T for product nuclei produced in deuteron induced reaction with 3.65 GeV/N incident energy. The values of α =B/2 are taken from Tab.1

| Mass region | $E^*_A$ | α=B/2 | ρ/ρ$_0$ | T (MeV) |
|---|---|---|---|---|
| 24≤A≤28 | 5.05 | 0.275 | 0.50 | 5.59 |
| 42≤A≤58 | 3.34 | 0.365 | 0.74 | 5.21 |
| 67≤A≤77 | 2.45 | 0.40 | 0.89 | 4.75 |
| 81≤A≤89 | 2.39 | 0.435 | 0.90 | 4.72 |
| 90≤A≤99 | 1.28 | 0.45 | 0.99 | 3.56 |
| 104≤A≤109 | 0.86 | 0.49 | 1 | 2.92 |

Since the data obtained in present work are inclusive, products in different mass regions may come from different mechanisms as we conclude above (from discussion of values of parameter B). This is also evident from the excitation energy, density and temperature values obtained with the catcher technique (see tables 5, 6). One can see from Tables 5,6 that the isoscaling parameter α increases with decrease of excitation energies and temperature T.

**5. Conclusion**

Isospin effects in $^{12}$C ion induced reactions on enriched tin isotopes are investigated. The values of fitting parameter B are determined for different mass regions of product nuclei. The discussion of obtained results show that parameter B is sensitive to the reaction mechanism: the light fragments are formed in the multifragmentation process, product nuclei in the intermediate mass regions (40≤A<60) are formed both in spallation and multifragmentation processes and heavy residuals (near of target mass) are formed via spallation only.

Besides, for reactions with $^{12}$C projectile the parameter B depends on the target neutron excess (ΔN$_t$) or the difference of asymmetry according to B=c+d·[(Z$_t$/A$_{1t}$)$^2$-(Z$_t$/A$_{2t}$)$^2$] equation. The value of B increases as the difference of asymmetry is increasing as it has been noticed in our early investigations in proton and deuteron induced reactions.

Using our early results of measurements of "thick-target thick-catcher" techniques we have determined the excitation energies per nucleon and temperatures of product nuclei in different mass regions, which allowed us to obtain density ratio ρ/ρ$_0$ for these mass regions. The values of excitation energy per nucleon, density and temperature confirm the conclusion about different mechanisms for formations of product nuclei. Besides, the coefficient of symmetry term γ has been determined for the light fragment, which is equal to 11.45 MeV and 10.48 MeV in proton and deuteron induced reactions, respectively.

**Acknowledgment.** We would like to express our gratitude to the operating personnel of the JINR Nuclotron for providing good beam parameters and to the first-year Master student A.G.Movsesyan for her help in preparing this paper.